\documentclass[a4paper,11pt]{article}
\pdfoutput=1
\usepackage{jcappub}
\usepackage[T1]{fontenc}

\usepackage{xspace}
\usepackage{subfig}
\usepackage{epstopdf}
\usepackage{units}
\usepackage{graphicx}
\usepackage{dcolumn}
\usepackage{bm}
\usepackage{epsfig}
\usepackage{url}
\usepackage{aas_macros}

\newcommand*{\los}{line of sight\xspace}
\newcommand*{\code}{\textsf{SONG}\xspace}
\newcommand*{\SN} {S/N\xspace}
\newcommand*{\F} [2] {\ensuremath{F^{{#1},{#2}}}\xspace}
\newcommand*{\lmax} {\ensuremath{\ell_\text{max}}\xspace}
\newcommand*{\Lmax} {\ensuremath{L_\text{max}}\xspace}

\renewcommand*{\O} [1] {\ensuremath{\mathcal{O}\left( {#1} \right)}}
\newcommand*{\fnlcon}{\ensuremath{f_{\text{NL}}^\text{intr}}\xspace}
\newcommand*{\fnl}{\ensuremath{f_{\text{NL}}}\xspace}
\newcommand*{\sci} [2] {\ensuremath{{#1} \times 10^{#2}}} 

\newcommand*{\eg} {e.\,g.\xspace}

\newcommand{\rmd}{\textrm{d}}

\renewcommand{\vec}[1]{{\mathbf {#1}}} 

\newcommand*{\tref} [1] {Table~\ref{#1}\xspace}

\newcommand*{\sref} [1] {Section~\ref{#1}\xspace}

\newcommand*{\eref} [1] {Eq.~\eqref{#1}\xspace}

\newcommand*{\fref} [1] {Figure~\ref{#1}\xspace}

\newcommand*{\cref} [1] {Chapter~\ref{#1}\xspace}

\title{The intrinsic bispectrum of the Cosmic Microwave Background}

\author{Guido W.\ Pettinari,}
\author{Christian Fidler,}
\author{Robert Crittenden,}
\author{Kazuya Koyama and}
\author{David Wands}
\affiliation{Institute of Cosmology and Gravitation,
University of Portsmouth,
Dennis Sciama Building,
Burnaby Road,
Portsmouth
PO1 3FX,
United Kingdom}
\emailAdd{Guido.Pettinari@port.ac.uk}
\emailAdd{Christian.Fidler@port.ac.uk}
\emailAdd{Robert.Crittenden@port.ac.uk}
\emailAdd{Kazuya.Koyama@port.ac.uk}
\emailAdd{David.Wands@port.ac.uk}

\abstract{We develop a new, efficient code for solving the second-order Einstein-Boltzmann equations, and use it to estimate the intrinsic CMB non-Gaussianity arising from the non-linear evolution of density perturbations.
The full calculation involves contributions from recombination and less tractable contributions from terms integrated along the \los.  We investigate the bias that this intrinsic bispectrum implies for searches of primordial non-Gaussianity. We find that the inclusion or omission of certain \los terms can make a large impact. When including all physical effects but lensing and time-delay, we find that the local-type \fnl would be biased by $\fnlcon=0.5$, below the expected sensitivity of the Planck satellite. The speed of our code allows us to confirm the robustness of our results with respect to a number of numerical parameters.}

\begin{document}
\maketitle
\flushbottom


\section{Introduction}
\label{sec:introduction}

Since cosmic inflation was proposed as a mechanism to solve the causality and flatness problems \cite{guth:1981a, linde:1982a, albrecht:1982a, starobinsky:1980a} and to generate primordial fluctuations \cite{hawking:1982a, starobinsky:1982a, mukhanov:1981a, bardeen:1983a}, many different theoretical models of inflation have been put forward. In most cases, it is difficult to distinguish between these various models just from the measurements of the power spectrum and it is useful to have more observables. Single-field slow-roll inflation produces an effectively Gaussian distribution of primordial adiabatic density perturbations \cite{maldacena:2003a, acquaviva:2003a}. Alternative models such as the curvaton scenario \cite{linde:1997a, enqvist:2002a,lyth:2002a,moroi:2001a,moroi:2002a}, can generate distinctive non-Gaussian features in the primordial fluctuations, which are potentially detectable in the bispectrum of the Cosmic Microwave Background (CMB). Such a detection would rule out the simplest models of inflation and strongly constrain the physics of the early Universe.

The bispectrum of the CMB anisotropies provides an optimal way to measure non-Gaussianity of primordial perturbations \cite{komatsu:2001a}. In models such as the curvaton one, where non-Gaussianity arises due to the non-linear evolution of the primordial curvature perturbation on super-horizon scales, the bispectrum peaks at squeezed configurations where one of the momenta is much smaller than the other two momenta. This is called the local type non-Gaussianity \cite{komatsu:2001a, gangui:1994a, verde:2000a} as the non-linearity appears locally in real space. On the other hand, the non-linearity of quantum fluctuations on sub-horizon scales during inflation generally produces a bispectrum that peaks for more equilateral configurations \cite{alishahiha:2004a, silverstein:2004a}. Theoretical templates for the bispectra have been developed to optimally measure these two distinct types of non-Gaussianity. In addition, an orthogonal template with minimal overlap was developed to measure the bispectrum that cannot be captured by the local and equilateral templates \cite{senatore:2010a}. These three templates have been applied to CMB anisotropies measured by WMAP, giving constraints $-3<\fnl^{local}<77$, $-221<\fnl^{eq}<323$, $-445 < \fnl^{orth} < -45$ at $95\%$ confidence level \cite{bennett:2012a}. The Planck satellite \cite{planck-tauber:2010a} will dramatically improve these constraints providing a possibility to distinguish between various early Universe models.

However, we do not expect all of the observed non-Gaussianity to be of primordial origin. Non-linear evolution will generate some degree of non-Gaussianity even in the absence of a primordial signal, for the simple reason that the product of Gaussian random fields is non-Gaussian. The propagation of CMB photons in an inhomogeneous Universe and their non-linear collisions with electrons make it possible for Gaussian initial conditions to be non-linearly propagated into a non-Gaussian temperature field. This results in the emergence of an \emph{intrinsic} CMB bispectrum, which is the topic of this paper.

The intrinsic CMB bispectrum acts as a systematic bias in the measurement of the primordial bispectrum \cite{komatsu:2010a}. In order to correctly interpret the upcoming non-Gaussianity measurements, it is of crucial importance to quantify this bias, which we label \fnlcon. In addition to that, the non-Gaussian signal from non-linear dynamics has an interest of its own, as it might shed light on the details of the gravity theory \cite{gao:2011a}.

The non-linear signal can be quantified theoretically by using second-order perturbation theory; this is the leading order of non-Gaussianity since linear evolution cannot generate non-Gaussian features that are not already present in the initial conditions. The Einstein and Boltzmann equations at second order have been studied in great detail \cite{bartolo:2006a, bartolo:2007a, pitrou:2007a, pitrou:2009b, beneke:2010a}. They are significantly more complicated than at first order and solving them numerically is a daunting task.

Existing approximations either neglect some of the physics or focus on a particular bispectrum configuration. On super-horizon scales at recombination, where only gravitational effects are important, it is well established that $\fnlcon \sim -1/6$ for the local model \cite{boubekeur:2009a, bartolo:2004a, bartolo:2004b}. On small angular scales, one has to consider the interactions taking place between photons and baryons before the time of decoupling. The contribution to \fnlcon arising from the fluctuations in the free-electron density has been shown to be of order unity \cite{senatore:2009a, khatri:2009a}, and likewise for the contribution from the other quadratic sources in Boltzmann equation \cite{nitta:2009a}. An alternative approach consists of focussing on the squeezed limit, where the local template peaks. The recombination bispectrum in this limit can be obtained by a coordinate rescaling \cite{creminelli:2004a} and yields a contamination to the local signal again of order unity \cite{creminelli:2004a, creminelli:2011a, bartolo:2012a, lewis:2012a}.

None of these estimates constitute a significant bias for Planck, which is expected to constrain the local model with an uncertainty of $\sigma_{f_{NL}} \sim 5$ \cite{komatsu:2001a}. However, the first full numerical computation of the bias, performed by Pitrou et al. (2010) \cite{pitrou:2010a, pitrou:2011a}, found the much higher value $\fnlcon \sim 5$, just at the detection threshold for Planck.

In this paper, we introduce a new code, \code (Second-Order Non-Gaussianity), to solve the second order Einstein-Boltzmann equations for photons, neutrinos, baryons and cold dark matter. The code is written in C, is parallel, and is based on the first-order Boltzmann code CLASS \cite{lesgourgues:2011a, blas:2011a}, from which inherits its modular structure and ease of use. \code is fast enough to perform various convergence tests to check the robustness of the numerical results. Utilising this code, we will study the intrinsic non-Gaussianity to quantify the bias in the measurements of primordial non-Gaussianity and evaluate its signal-to-noise ratio.


While this paper was in preparation, two papers appeared that study the intrinsic bispectrum, giving similar results for the bias to the primordial non-Gaussianity templates, but different ones for the signal-to-noise ratio \cite{huang:2012a, su:2012a}. We will discuss later why Refs.~\cite{pitrou:2010a, huang:2012a, su:2012a} obtained different results.


This paper is organised as follows. In \sref{sec:boltzmann}, we present the second order Boltzmann equation and explain physical meanings of various terms. In \sref{sec:differential}, we explain how we solve Einstein and Boltzmann equations using our code, \code. In \sref{sec:line_of_sight}, the \los integration is introduced in order to solve Boltzmann equation up to today. We will explain which contributions in the \los integration are included in our analysis. In \sref{sec:bispectrum}, we define the bispectrum of temperature anisotropies and introduce the Fisher matrix to estimate the importance of the intrinsic bispectrum.
Section~\ref{sec:results} is devoted to present our main results, while in \sref{sec:convergence} we show their numerical robustness. Finally, we discuss our results in \sref{sec:conclusions}.


\paragraph{Metric}
We employ the conformal Poisson gauge \cite{bertschinger:1995a, bruni:1997a}, where the metric reads:
\begin{equation}
  \rmd s^2 = a^2(\tau) \, \left\{ - (1+2\Psi) \rmd \tau^2
    + 2 \omega_i \rmd x^i \rmd \tau
  + \,\left[\,(1-2\Phi) \delta_{ij} + 2\,\gamma_{ij} \,\right]\, \rmd x^i \rmd x^j \right\} \;.
\label{eq:metric}
\end{equation}
The vector potential is transverse $\omega_{i, i} = 0$, while the spatial metric is both transverse $\gamma_{ij,i} = 0 $ and traceless $\gamma_{ii} = 0$.

\paragraph{Initial conditions}
We shall assume scalar adiabatic initial conditions which are completely Gaussian in the primordial curvature perturbation $\zeta$. 


\paragraph{Cosmological parameters}
Thorough the paper we shall employ a $\Lambda$CDM model with WMAP7 parameters \cite{komatsu:2011a}, whereby $h = 0.704$, $\Omega_b = 0.0455$, $\Omega_{\text{cdm}} = 0.228$, $\Omega_\Lambda = 0.727$, $A_s = \sci{2.43}{-9}$, $n_s = 0.967$, $\tau_\text{reio} = 0.088$, $N_\text{eff} = 3.04$. In this model, the age of the Universe is $\unit[13.73]{Gyr}$, the conformal age $c\,\tau_0 = \unit[14329]{Mpc}$ and recombination happens at $z=1088$, corresponding to a conformal time of $c\,\tau_\text{rec} = \unit[284]{Mpc}$.


\paragraph{Second order perturbations}
We solve the Einstein and Boltzmann equations at second order in the cosmological perturbations. We expand every physical quantity according to $X \,\simeq\, X^{(0)} + X^{(1)} + X^{(2)}$, so that a second-order equation has the same structure of the corresponding first-order one, plus a source term that contains terms quadratic in the first-order perturbations. We assume that the vector and tensor modes of the metric are generated only by second-order effects, which implies $\omega_i^{(1)}=0$ and $\gamma_{ij}^{(1)}=0$.
\section{Boltzmann equation}
\label{sec:boltzmann}

The time evolution of the photon distribution function $f(\eta, \vec{x}, q\vec{n})$, with $q$ being the comoving momentum, is determined by the Boltzmann equation
\begin{equation}
	\label{eq:boltz}
	\frac{\partial f}{\partial \eta} + \frac{dx^i}{d\eta}\frac{\partial f}{\partial x_i} + \frac{dq}{d\eta}\frac{\partial f}{\partial q}  + \frac{dn^i}{d\eta}\frac{\partial f}{\partial n_i} \,= \,\mathfrak{C}[f] \;,
\end{equation}
where the left-hand side describes particle propagation in an inhomogenous space-time and the right-hand side describes the interactions between photons and electrons through Compton scattering. The second term encodes free streaming, that is the propagation of perturbations from the small to the large multipoles. At higher order this term also includes time delay effects. The third term, at background level, causes the redshifting of photons, and at higher-order includes the well-known Sachs-Wolfe (SW), integrated Sachs-Wolfe (ISW) and Rees-Sciama (RS) effects. The fourth term vanishes to first order and describes the small-scale effect of gravitational lensing on the CMB.

It is convenient to use the brightness $\Delta$, defined as the momentum-integrated distribution function normalised to the background energy density
\begin{equation}
	\label{eq:Deltadef}
	1+\Delta(\eta,\vec{x},\vec{n}) = 
	\frac{\int dq \, q^3 \, f(\eta,\vec{x},q\vec{n})}
	{\int dq \, q^3 \, f^{(0)}(q)} \;,
\end{equation} 
where $f^{(0)}$ is the background distribution function, which takes the form of a black-body distribution.
The terms in the Boltzmann equation (\ref{eq:boltz}) are explicitly given, up to second order, by \cite{beneke:2010a, pitrou:2009a}:
\begin{align}\nonumber
	&\frac{dx^i}{d\eta}\frac{\partial f}{\partial x_i} &\rightarrow& \quad\quad  (1 + \Psi + \Phi)\,n^i\partial_i\Delta \;, \\ \nonumber
	&\frac{dq}{d\eta}\frac{\partial f}{\partial q} &\rightarrow& \quad\quad 4 \, (n^i\partial_i \Psi -\dot{\Phi} - n^i \dot{\omega}_i + n^i n^j \dot{\gamma}_{ij})\,(1 + \Delta) -4 (\Psi -\Phi)\,n^i\partial_i \Psi -8 \Phi\dot{\Phi} \;, \\
	&\frac{dn^i}{d\eta}\frac{\partial f}{\partial n_i} &\rightarrow& \quad\quad -(\delta^{ij} - n^i n^j)\,\partial_j(\Psi+\Phi)\,\frac{\partial \Delta}{\partial n^i}
 \,.
	\label{eq:timedelay_lensing_redshift}
\end{align}
We represent the terms containing only metric perturbations by 
\begin{equation}
\mathcal{M} \,\,=\,\,  4 \, (n^i\partial_i \Psi -\dot{\Phi} - n^i \dot{\omega}_i + n^i n^j \dot{\gamma}_{ij}) \,-\, 4 (\Psi -\Phi)\,n^i\partial_i \Psi \,-\, 8 \Phi\dot{\Phi}
\,,
\end{equation}
and the quadratic terms that include the photon distribution $\Delta$ by 
\begin{equation}
 \label{eq:QL}
\mathcal{Q}^L \,\,=\,\, (\Psi + \Phi)\,n^i\partial_i\Delta \,\,+\,\, 4 \, (n^i\partial_i \Psi -\dot{\Phi})\,\Delta \,\,-\,\, (\delta^{ij} - n^i n^j)\,\partial_j(\Psi+\Phi)\,\frac{\partial \Delta}{\partial n^i}
\,.
\end{equation}
In the following, we shall refer to the parts of $\mathcal{Q}^L$ as time-delay (first term), redshift (second term) and lensing (third term). 
Note that the above expressions are valid only when $\Delta$ and the metric variables are expanded up to second order and when $\omega_i^{(1)}=0$ and $\gamma_{ij}^{(1)}=0\,$.

We perform a Fourier transform from real space, ${\vec x}$, into comoving wavevector, $\vec{k}$, and transform from photon direction, $\vec{n}$, into spherical harmonics, $(\ell,m)$. 
This transforms the Boltzmann and Einstein equations from partial differential equations into hierarchies of ordinary differential equations. 
We will use a single composite index, $n$, to express the harmonic dependence,  $(\ell,m)$, and polarisation. The Boltzmann equation (\ref{eq:boltz}) then reads
\begin{equation}
	\label{eq:Boltzmann}
	\dot{\Delta}_{n} + k\,\Sigma_{nm}\Delta_{m} + \mathcal{M}_{n}  +\mathcal{Q}^L_{n} \;=\; \mathfrak{C}_n \;,
\end{equation}
where $\Sigma_{nm}$ is the free streaming matrix that arises from the decomposition of $n^i\partial_i\Delta$ into spherical harmonics. 
The collision term
\begin{equation}
	\label{eq:collision_term}
	\mathfrak{C}_n = -|\dot{\kappa}|\,\left(\;\Delta_{n} - \Gamma_{nm}\Delta_{m}-\mathcal{Q}^C_{n}\,\,\right) \;,
\end{equation}
includes the Compton scattering rate $|\dot{\kappa}|$ and consists of three distinct contributions: the purely second-order loss term $-|\dot{\kappa}|\,\Delta_{n}$, describing scatterings out of a given mode $\Delta_n$, and scatterings into that mode from purely second-order contributions, $|\dot{\kappa}|\,\Gamma_{nm}\Delta_{m}$, and quadratic contributions, $|\dot{\kappa}|\,\mathcal{Q}^C_n$. When the scattering rate is large
photons and baryons behave as a tightly coupled fluid, where all moments larger than the quadrupole are suppressed.

\section{Differential system}
\label{sec:differential}

At linear order Fourier modes evolve independently. At second order the presence of quadratic source terms leads to mode coupling. In Fourier space this corresponds to a convolution between two wavemodes $\vec{k_1}$ and $\vec{k_2}$. Accordingly, we express a second-order perturbation $\Delta^{(2)}$ as a convolution of its \emph{transfer function} $\mathcal{T}_\Delta^{(2)}$ with a quadratic product of the primordial curvature potential $\phi$:
\begin{equation}
	\label{eq:transfer}
	\Delta^{(2)}(\vec{k}, \vec{n}) \, \equiv \, \int \frac{\text{d}\vec{k_1} \, \text{d}\vec{k_2}}{(2\pi)^3} \, \delta (\vec{k_1} + \vec{k_2} + \vec{k}) \,
	\mathcal{T}_\Delta^{(2)}\,(\vec{k},\vec{k_1},\vec{k_2}, \vec{n})\,
	\phi(\vec{k_1})\,\phi(\vec{k_2})
\,.
\end{equation}
The main advantage of this form is that the stochastic potential is clearly separated from the deterministic transfer function, whose evolution is determined by the Boltzmann and Einstein equations.
However, the transfer functions depend on two wavemodes $\vec{k_1}$ and $\vec{k_2}$, which, considering a homogeneous and isotropic spacetime, reduce to three degrees of freedom. We choose to use the magnitudes of the three wavevectors ($k_1$, $k_2$ and $k=|\vec{k_1}+\vec{k_2}|$) and solve the Einstein-Boltzmann system of differential equations on a 3D-grid. Note that the limits of the third direction, $k$, are dictated by the triangular condition in the Dirac-delta function in \eref{eq:transfer}, so that our k-space is a tetrahedron rather than a cube.

Our numerical code \code solves Boltzmann equations for photons, massless neutrinos, baryons and cold dark matter, and includes perturbed recombination as described in Ref.~\cite{senatore:2009b}. For the photons, we consider temperature, E- and B-mode polarisation. We evolve the perturbations for the relativistic species until after recombination, including all terms in \eref{eq:Boltzmann} and \eref{eq:timedelay_lensing_redshift}, and obtain their value today by then employing the \los formalism, which we describe in \sref{sec:line_of_sight}.
We truncate the relativistic hierarchies at $\ell\sim50$, where we reach convergence for recombination effects. In a typical run of our code, we evolve a system of $\sim 100$ differential equations for $\sim 10^6$ independent $(k_1,k_2,k)$ triplets. By using \emph{ndf15}, the implicit differential solver of the first-order Boltzmann code \emph{CLASS} \citep{lesgourgues:2011a, blas:2011a}, we manage to perform this task in $\sim 1$ hour on a quad-core machine.


Our initial conditions are set in the radiation dominated era when all Fourier modes are super-horizon and in the tightly coupled regime. Our results are then not sensitive to the precise starting time of integration. 

We have compared our second-order transfer functions with several known analytical limits. We find a $<1\%$ agreement with the sub-horizon kernels of standard perturbation theory \cite{bernardeau:2002a, goroff:1986a, makino:1992a, jain:1994a}. We obtain similar agreement with the squeezed-limit transfer functions derived in Refs.~\cite{creminelli:2011a, bartolo:2012a} and with the numerical transfer functions produced by \emph{CMBquick} \cite{pitrou:2010a, pitrou:2011a}. Furthermore, our code is stable with respect to the choice of the Einstein equation used to evolve the curvature potential. 

Finally, we have compared our results with an updated version of a code based on Green's functions rather than transfer functions \cite{beneke:2011a}. Green's functions provide an orthogonal method of reducing the stochastic Boltzmann equations to algebraic differential equations, that can be solved efficiently. The Green's function $G_{nm}(k,\eta_1,\eta_2)$ depends on two times and describes the impact of a mode $m$ at time $\eta_2$ on the mode $n$ at time $\eta_1$. The differential equations for the Green's functions are especially simple as they are independent of the quadratic source terms. It is also not necessary to introduce additional wavevectors $k_1$ and $k_2$. However, the Green's functions do depend on an additional time, $\eta_2$, and have one additional composite index $m$. For runs with average precision, the methods have a comparable speed, but, when refining the numerical parameters, we find that the transfer function approach scales better. Comparing the results between these different approaches, we obtain a sub-percent level agreement.

\section{Line of Sight}
\label{sec:line_of_sight}

After recombination, photons stream so that higher multipoles with $l\approx k (\eta-\eta_{\text{rec}})$ are excited. The Boltzmann hierarchy can be no longer truncated at low multipoles, making the numerical treatment computationally inefficient. Instead we will compute the moments today, $\eta_0$, using the \los integration
\begin{equation}
	\label{eq:los_integral}
	\Delta_{n}(\eta_0) 
	= \int_0^{\eta_0}  d\eta \,e^{-\kappa(\eta)} j_{nm}(k(\eta_0-\eta))\,\mathcal{S}_m(\eta) \;,
\end{equation}
which is an integral representation for the solution of the differential equation \cite{seljak:1996a}
\begin{equation} \label{eq:Losdgl}
	\dot{\Delta}_{n} + k \, \Sigma_{nm}\Delta_{m} = -|\dot{\kappa}|\Delta_{n} 
	+\mathcal{S}_n \;.
\end{equation}
The excitation of higher multipoles through streaming is encoded in the projection functions $j_{nm}$, which are linear combinations of spherical Bessel functions \cite{beneke:2011a}.
The source $\mathcal{S}_{n}$ is obtained by comparing equation (\ref{eq:Losdgl}) with the full Boltzmann equation (\ref{eq:Boltzmann}):
\begin{equation}
	\label{eq:line_of_sight_sources}
	\mathcal{S}_n =-\mathcal{M}_{n} -\mathcal{Q}^L_{n}+ |\dot{\kappa}|\left(\Gamma_{nm}\Delta_{m}+\mathcal{Q}^C_{n}\right) \;.
\end{equation}
%

The source terms including $|\dot{\kappa}|$ are relevant only on the last scattering surface (LSS), where the visibility function $|\dot{\kappa}|e^{-\kappa}$ peaks. These are computed by evolving the full differential system through recombination, as described in the previous section. Thereafter we may safely neglect them when computing the \los integral (\ref{eq:los_integral}) for $\eta\gg\eta_{LSS}$.

On the other hand, the metric term $\mathcal{M}_n$ and the quadratic terms $\mathcal{Q}^L_n$ are important at all times. The metric contributions can be computed up to today evolving a simplified Boltzmann hierarchy as the radiation component of the Universe becomes quickly negligible after matter-radiation equality. The contribution $\mathcal{Q}^L_n$ is purely quadratic in first order terms and, in principle, can be computed without the need to solve the differential system at second order. However, it is impractical to include it numerically in a standard \los approach as it comprises a sum over multipoles which are important over all scales and times.

\subsection{Treating the redshift contribution}
\label{subs:delta_tilde}

As shown by Huang and Vernizzi \cite{huang:2012a}, the redshift contribution to $\mathcal{Q}^L_n$ in \eref{eq:QL}
\begin{equation}
4\, \,(n^i\partial_i \Psi \,-\, \dot{\Phi})\,\Delta
\end{equation}
can be absorbed by using the new variable
\begin{equation}
	\label{eq:transformation}
 	\tilde{\Delta} \,\equiv\, \ln \,(1 + \Delta) \,. 
\end{equation}
%
The time derivative of $\tilde\Delta$ up to second order is then given by
\begin{eqnarray}
	\nonumber
	\dot{\tilde{\Delta}} = \dot{\Delta} - \Delta\dot{\Delta} &=&- n_i \partial^i\tilde{\Delta} \,-\, \mathcal{M} \,+\,\mathfrak{C}(1-\Delta) \\[0.15cm]
	&& -\mathcal{Q}^L \,+\, 4 \, (n^i\partial_i \Psi -\dot{\Phi})\Delta \;,
  \label{eq:absorption}
\end{eqnarray}
where we have used the first-order Boltzmann equation
\begin{equation}
	\dot{\Delta} = -n_i\partial^i \Delta -4 \, (n^i\partial_i \Psi -\dot{\Phi}) +\mathfrak{C}  \,,
\end{equation}
to replace the quadratic term $\Delta\dot{\Delta}$ and the second-order one \eref{eq:Boltzmann} to replace $\dot{\Delta}$.
The new contribution $4\, (n^i\partial_i \Psi -\dot{\Phi} )\Delta$ exactly cancels the redshift term in $\mathcal{Q}^L$, so that the second line of \eref{eq:absorption} reduces to only the time-delay and lensing contributions. In addition, the collision term $\mathfrak{C}$ is replaced by $\mathfrak{C}\,(1-\Delta)$\,.

As can be seen, the transformation is effective because the second order source we are eliminating is the first order $\Delta$ times part of the first order source.  The price is to make the scattering term more complex.  However the change is not localised to the LSS; like $\Delta$, $\tilde\Delta$ is non-linearly related to the observed temperature anisotropies. This leads to an additional quadratic contribution to the temperature bispectrum arising from the first-order evolution, as we show in \sref{sec:reduced_bispectrum}. 

Unfortunately, this still leaves other problematic terms in $\mathcal{Q}^L$, the lensing and time-delay terms.  These do not relate to the first-order sources, and cannot be removed by a similar change of variables.  We will not include them in the \los integration in this paper, and leave them for future work.  Note, however, that we do include all terms in $\mathcal{Q}^L$ when solving the differential system given by \eref{eq:Boltzmann} up to recombination.

%

\subsection{A note on integration by parts}
\label{subs:integration_by_parts}

It is often a good technique to use integration by parts in order to separate recombination effects from time-integrated effects. By doing so, \eref{eq:los_integral} becomes
\begin{eqnarray}
	\int_0^{\eta_0} \, d\eta \,e^{-\kappa}\,\mathcal{S}_m \,\, j_{nm}(k(\eta_0-\eta))\, 
	=
	\int_0^{\eta_0} \, d\eta \,e^{-\kappa} \,\left(\frac{\dot{\mathcal{S}}_m}{k} - \frac{\dot{\kappa}\mathcal{S}_m}{k}\right) \, J_{nm}(k(\eta_0-\eta))
	\;,
\end{eqnarray}
where we have chosen to integrate $j_{nm}$ and $J_{nm}$, its antiderivative, can still be expressed in terms of spherical Bessel functions. This is usually done at first order, where the source is equal to the gradient of the potential $\mathcal{S}_m = k_m\,\Phi $ and gives rise to the usual SW ($\dot\kappa \Phi$) and ISW ($\dot\Phi$) split. This separation is useful because $\dot\Phi$ is much smaller than $k\Phi$ as the potential is slowly changing. The second-order metric terms $\mathcal{M}$ can be treated in the same way.

However, the quadratic sources $\mathcal{Q}^L$ are problematic as they contain the first-order photon fluctuations, which oscillate with frequency $k$ so that $k^{-1} \dot{\mathcal{Q}^L}_n \sim \mathcal{Q}^L_n$. Integration by parts then generates two terms: one with $|\dot\kappa|$, which is clearly located on the LSS, and a second one which is comparable to the original integral. That second term itself can be decomposed by using integration by parts, and will yield a non-negligible LSS contribution. Therefore, the technique fails to single out a unique LSS contribution.

When we exclude sources such as lensing, we exclude them in their entirety rather than imposing an arbitrary split. In this way, our results can be complemented by the known non-perturbative approaches, see Ref.~ \cite{lewis:2012a, hanson:2009a, smith:2011a, serra:2008a, lewis:2011a, lewis:2006a} for lensing.

\section{Temperature Bispectrum}
\label{sec:bispectrum}

As described in the previous sections, we are able to efficiently compute the full second-order transfer functions. One of the most interesting quantities that may be derived using these transfer functions is the intrinsic non-Gaussianity, which is usually quantified by computing the three-point correlation, or bispectrum. 
In contrast to the angular power spectrum $C_{\ell}$, which only depends on the angular separation of two points, the bispectrum can be computed for all possible triangles in the CMB sky, resulting in three angular separations or three multipole indices for the bispectrum.

\subsection{The observed temperature}
We first relate the brightness $\Delta$ to the observed temperature perturbations.
It is not generally possible to define a temperature in a perturbed Universe, because the notion of temperature implies thermal equilibrium. Perturbations provoke an unbalanced transfer of momentum between photons and baryons that breaks the blackbody spectrum of the photon distribution function. While at first order the blackbody shape of the spectrum is preserved \cite{dodelson:2003b}, at second order this is no longer the case \cite{pitrou:2010b}. Due to spectral distortions, the effective ``temperature'' perturbation $\Theta$ becomes momentum dependent, $\Theta(\eta,\vec{x},\vec{q})$; this is defined as 
\begin{equation}
	f(\eta,\vec{x},\vec{q}) = \left[\exp\left(\frac{q}
	{T \,(1+\Theta(\eta,\vec{x},\vec{q}))}\right)-1\right]^{-1}\;,
\end{equation}
where $T$ is the background temperature. When we neglect spectral distortions, the above equation, together with the definition \eref{eq:Deltadef}, allows us to define the bolometric temperature perturbation as
\begin{equation}
	\label{eq:bolometric_temperature}
	1+\Delta = (1+\Theta)^4\;,
\end{equation}
which is expanded up to second order into
\begin{equation}
	\label{eq:delta_bolometric}
	\Delta \:=\: 4\,\Theta + 6\,\Theta \, \Theta	\;,
\end{equation}
or equivalently, using \eref{eq:transformation},
\begin{equation}
	\label{eq:delta_tilde_bolometric}
	\tilde\Delta \:=\: 4\,\Theta - 2\,\Theta \, \Theta \;.
\end{equation}
It was shown in Ref.\ \cite{pitrou:2010b} that neglecting spectral distortions is indeed a good approximation for computing bispectra. 

\subsection{Reduced Bispectrum}
\label{sec:reduced_bispectrum}

The temperature multipoles $a_{\ell m}$ are defined by %
\footnote{The extra $\ell$-dependent factors in the integral counter the factors included in the multipole decomposition of $\Theta_{\ell m}$ in order to simplify the Boltzmann equations.}
\begin{equation}
a_{\ell m} = \int d\Omega \, Y^*_{\ell m} \, \Theta(\vec{n}) = \int \frac{\text{d}\vec{k}}{(2\pi)^3}\,(-i)^{l}\sqrt{\frac{4\pi}{2l+1}}\Theta_{\ell m}(\vec{k})\;.
\end{equation}
The temperature bispectrum is then given by the three-point correlation of the multipoles, 
\begin{eqnarray} \nonumber \label{eq:bispectrum_full}
\langle a_{\ell_1 m_1} a_{\ell_2 m_2} a_{\ell_3 m_3}\rangle&=& (-i)^{\ell_1 + \ell_2 + \ell_3}\sqrt{\frac{(4\pi)^3}{(2\ell_1+1)(2\ell_2+1)(2\ell_3+1)}}\\[0.2cm]
&&\int\frac{\text{d}\vec{k}_1\text{d}\vec{k}_2\text{d}\vec{k}_3}{(2\pi)^9}\left\langle \Theta_{\ell_1 m_1}(\vec{k}_1) \Theta_{\ell_2 m_2}(\vec{k}_2) \Theta_{\ell_3 m_3}(\vec{k}_3)\right\rangle.
\end{eqnarray}
Statistical isotropy enforces strict bounds on the multipole structure of this object. The multipole indices are dictated by the Gaunt structure allowing us to define the reduced temperature bispectrum $b_{\ell_1 \ell_2 \ell_3}[\Theta]$
\begin{equation}
 \langle a_{\ell_1 m_1} a_{\ell_2 m_2} a_{\ell_3 m_3}\rangle \,=\, \mathcal{G}_{\ell_1\,\ell_2\,\ell_3}^{m_1 m_2 m_3} \,b_{\ell_1 \ell_2 \ell_3} [\Theta]\;.
\end{equation}

Using the relations \eref{eq:delta_bolometric} and \eref{eq:delta_tilde_bolometric} we can relate the temperature bispectrum to the analogous bispectra constructed using the brightness moments $\Delta$ and $\tilde\Delta$, which are computed by our code, by:
\begin{eqnarray} \nonumber
b_{\ell_1 \ell_2 \ell_3} [\Theta] &=& \,\frac{1}{4^3} \,\, b_{\ell_1 \ell_2 \ell_3}  [\Delta] -3 \, C_{\ell_1} C_{\ell_2} -3 \, C_{\ell_1} C_{\ell_3} -3 \, C_{\ell_2} C_{\ell_3}\\[0.1cm]
 &=& \,\frac{1}{4^3} \,\, b_{\ell_1 \ell_2 \ell_3}{[\tilde{\Delta}]} +C_{\ell_1} C_{\ell_2} +C_{\ell_1} C_{\ell_3}+ C_{\ell_2} C_{\ell_3} \;,
\label{eq:deltatilde_to_temp}
\end{eqnarray}
where the angular power spectrum of temperature fluctuations, $C_l$, is obtained from linear perturbation theory as $\langle a_{\ell m} a_{\ell' m'}\rangle \,=\, (-1)^m\,C_\ell\,\delta_{\ell\ell'}\,\delta_{m-m'}$.
In principle, the temperature bispectrum can be obtained by either computing $b_{\ell_1 \ell_2 \ell_3} [\Delta]$ or $b_{\ell_1 \ell_2 \ell_3} [\tilde\Delta]$. In practice, as explained in \sref{subs:delta_tilde}, using the latter is advantageous because the $\tilde\Delta$ variable includes by construction the numerically challenging redshift contribution.


\subsection{Bispectrum computation}
The full-sky bispectrum \eref{eq:bispectrum_full} is given by a multi-dimensional integration over $\vec{k_1}$, $\vec{k_2}$, $\vec{k_3}$. As it contains the photon perturbations at the present time, it is highly oscillatory in all three wavevectors. The leading contribution to the bispectrum combines one second order perturbation with two first order ones.

We choose our coordinate system so that the wavevector of the second-order transfer function, $\vec{k}$, is aligned with the zenith of the spherical harmonics. In this system, scalar contributions correspond to azimuthal number $m=0$. At second order, mode-coupling will generate all moments $m\neq 0$ even with scalar initial conditions. Any bispectrum $b_{\ell_1 \ell_2 \ell_3}$ can be split into these separate contributions:
\begin{equation}
b_{\ell_1 \ell_2 \ell_3} = \sum \limits_m b_{\ell_1 \ell_2 \ell_3}^{\{m\}}\;.
\end{equation}
In the squeezed case, where the large wave-vectors are effectively aligned with the zenith, the scalar contributions are dominant and the sum can be truncated at $m=0$. For non-squeezed configurations, the scalar bispectrum will still give an important contribution to the sum, but higher moments can no longer be neglected. In this work we focus on the scalar contributions and compute their impact on the bispectrum. 

We separate the statistical perturbations in \eref{eq:bispectrum_full} using the transfer functions defined in \eref{eq:transfer}. Then the statistical average only affects the primordial perturbations $\phi$ and it can be expressed using the primordial power spectrum $P(k)$, assuming that there is no primordial non-Gaussianity. The angular integrations can be solved analytically and we are left with four integrations over the magnitudes $(k,k_1,k_2)$ of the second order transfer function and one parameter $r$ of dimension time, which is introduced by the Rayleigh expansion of the three-dimensional Dirac-delta function:
\begin{eqnarray}
	\label{eq:bispectrum_integral}
	b\,_{\ell_1 \ell_2 \ell_3}^{\{0\}} \, = \, \left(\frac{2}{\,\pi}\right)^3 && \int dr\, r^2\, \int dk\,k^2\,\mathcal{T}_{\,l_30}^{(2)}(k,k_1,k_2)\,j_{l_3}(r k)\\[0.15cm]
	\nonumber
	&& \int dk_1\,k_1^2 \, \mathcal{T}_{\,l_1}^{(1)}(k_1)P(k_1)\,j_{l_1}(r k_1) \; \int dk_2\,k_2^2\,\mathcal{T}_{\,l_2}^{(1)}(k_2)P(k_2)\,j_{l_2}(r k_2) \, + \;\text{2 symm.}
\end{eqnarray}
The above equation is similar to the usual formula for the primordial bispectrum (see, \eg, Refs. \cite{komatsu:2001a, fergusson:2007a}), but with one crucial difference: the shape of the intrinsic second-order non-Gaussianity is not separable, meaning that the integration cannot be split into three one-dimensional integrations. We obviate this problem by using the fact that $\mathcal{T}^{(2)}(k,k_1,k_2)$ is smooth in $k_1$ and $k_2$ as a consequence of the \los integral acting only on $k$. After integrating over the highly oscillatory direction $k$, we interpolate the result in the smooth directions. Finally, we perform the remaining two oscillatory integrations using the same technique, effectively breaking the three-dimensional oscillatory integration into three one dimensional integrations.

We are able to compute the bispectrum to percent accuracy for $(100)^3$ configurations of $(l_1,l_2,l_3)$ in a few hours on a quad-core machine. This, together with the convergence tests in \sref{sec:convergence}, demonstrates that our implementation of this integration is fast and numerically stable.

\subsection{Squeezed limit}
\label{sec:squeezed_analytical_approximation}

For squeezed triangles, where the small-$k$ side is within the horizon today but was not at recombination, the intrinsic bispectrum is known  approximately \cite{creminelli:2004a}. In this configuration, the long-wavelength mode acts as a perturbation of the background that alters the observed angular scale of the  short wavelength modes. The reduced bispectrum for the bolometric temperature then takes the following form \cite{lewis:2012a,bartolo:2012a,creminelli:2011a}:
\begin{equation}
	\label{eq:analytical}
	b_{\ell_1 \ell_2 \ell_3} [\Theta]=  C_{\ell_1}C_{\ell_2} + C_{\ell_1}C_{\ell_3} + C_{\ell_2}C_{\ell_3}
									 - C_{\ell_1}^{T\zeta} \frac{1}{2} \left(%
	   							 	 C_{\ell_2} \frac{d\text{ln}\,(l_2^2\,C_{\ell_2})}{d\text{ln}\,l_2} +%
	   							   C_{\ell_3} \frac{d\text{ln}\,(l_3^2\,C_{\ell_3})}{d\text{ln}\,l_3}%
										 \right) \;,
\end{equation}
where $C_{\ell_1}^{T\zeta}$ is the correlation between the photon temperature and the super-horizon curvature perturbation $\zeta = \Delta/4 - \Phi$ at first order. The derivative term encodes the shift in the observed angular scales, known as Ricci focussing \cite{lewis:2012a}, while the first three terms represent the smaller effect due to anisotropic redshifting, known as redshift modulation \cite{lewis:2012a}. A quick comparison with \eref{eq:deltatilde_to_temp} shows that the bispectrum induced by Ricci focussing corresponds to the bispectrum of $\tilde\Delta$.

\begin{figure}[t]
	\centering
		\includegraphics[width=0.7\linewidth]{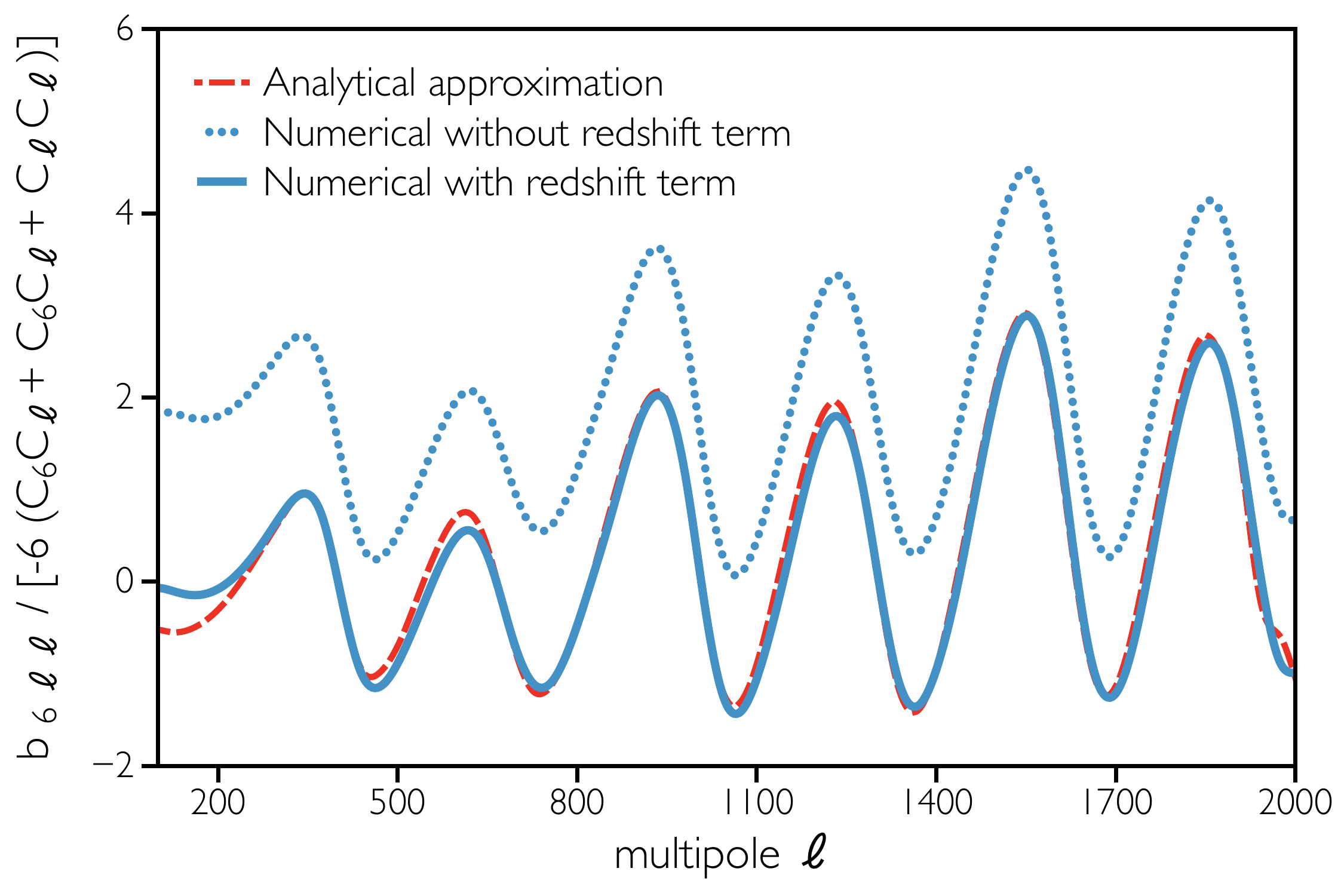}
	\caption{Numerical temperature bispectra versus the squeezed-limit approximation in \eref{eq:analytical} for a WMAP7 cosmology, where $\ell_1=6$ and $\ell_2=\ell_3=\ell\,$.  We normalise the curves with respect to the ultra-squeezed limit for a local-type bispectrum with $\fnl=1$ \cite{gangui:1994a, komatsu:2001a}, so that the primordial curve would appear as a constant horizontal line with amplitude close to unity.
	}
	\label{fig:analytical}
\end{figure}

In \fref{fig:analytical} we show two temperature bispectra obtained with \code compared to the analytical approximation for a squeezed configuration where the large-scale mode is fixed. The bispectrum computed using $\tilde\Delta$ (labelled $B^{R+Z}$ in \sref{sec:results}), which includes both the scattering sources and the time-integrated effect arising from the redshift term, matches the analytical curve to a precision of a few percent. On the other hand, the bispectrum computed using the standard brightness $\Delta$ (labelled $B^{R}$ in \sref{sec:results}), which does not include the redshift term, presents a nearly constant positive offset with respect to the analytical approximation.

\subsection{The estimator}
\label{subs:estimator}

We quantify the importance of the intrinsic bispectrum by using a Fisher matrix approach. The Fisher matrix element between two temperature bispectra $B^{\,i}$ and $B^{\,j}$ is given by \cite{komatsu:2001a}:
\begin{equation}
	\label{eq:estimator}
	\F{\,i}{j} \, = \, \sum \limits_{2\leq l_1\leq l_2\leq l_3}^{\lmax} \, \frac{B^{\,i}_{\,\ell_1 \ell_2 \ell_3}\,B^{\,j}_{\,\ell_1 \ell_2 \ell_3}}{C_{\ell_1}C_{\ell_2}C_{\ell_3}\,\Delta_{\ell_1 \ell_2 \ell_3}} \;,
\end{equation}
where the angular power spectrum of temperature fluctuations, $C_l$, is obtained from linear perturbation theory, $\Delta_{\ell_1 \ell_2 \ell_3} = 1,2,6$ for triangles with no, two or three equal sides, and \lmax is the maximum angular resolution attainable with the considered CMB survey.
The observability of the intrinsic bispectrum $B$ is quantified by its signal-to-noise: $\SN = \sqrt{\F{B}{B}} $, while the bias that it induces on the \fnl measurement of a primordial template T is given by $\fnlcon = \F{B}{T}/\F{T}{T}$. In what follows, we shall assume an ideal experiment without beam or noise, going up to $\lmax=2000$.

We obtain the Fisher matrix elements in \eref{eq:estimator} by interpolating the bispectra on a mesh \cite{pettinari:2013c}. This allows us to compute the numerically expensive second-order bispectrum for a small number of configurations, typically \O{100} per $\ell$-direction, and still estimate the sum in \eref{eq:estimator} to high accuracy.  This is a huge speed improvement over existing techniques, even when considering the simple separable bispectra. As an example, we can compute the signal-to-noise of the equilateral model for a given cosmology with $\sim 1\%$ accuracy in the matter of seconds on a quad-core machine (see \sref{sec:convergence}).

\section{Results}
\label{sec:results}

We present results for the intrinsic bispectrum considering three different combinations of \los sources. The first considered bispectrum ($B^R$) includes only sources located on the surface of last scattering, that is the $|\dot\kappa|$ sources in \eref{eq:line_of_sight_sources} plus the second-order Sachs-Wolfe effect, $4\,|\dot{\kappa}|\,\Psi$, which only contributes to the monopole. The second ($B^{R+Z}$) also includes the redshift term of $\mathcal{Q}^L$, that is $4 \, (n^i\partial_i \Psi -\dot{\Phi})\,\Delta$. This is computed using $\Tilde\Delta$ and it is the same bispectrum presented in Huang and Vernizzi (2012) \cite{huang:2012a}. Finally, $B^{R+Z+M}$ consists of the above sources plus all the terms in $\mathcal{M}$. One of such terms gives rise to the Rees-Sciama effect \cite{rees:1968a, boubekeur:2009a, mollerach:1995a, munshi:1995a}, which is given by $4 \, (\dot\Psi+\dot\Phi)$.  The latter bispectrum contains all terms in the Boltzmann equation but the time-delay and lensing contributions, and is therefore our most complete bispectrum.

 
We compute the contamination $\fnlcon$ induced by the intrinsic bispectra for three models of primordial non-Gaussianity: local, equilateral and orthogonal. Our results are shown in \tref{tab:fnl_results}, where we assume an ideal experiment with $\lmax = 2000$.

\begin{table}[htbp]
	\centering
   \begin{tabular}{l || *{3}c || r}
       Model       & ~~~~~~~$B^R$~~~~~~~~ & ~~~~~$B^{R+Z}$~~~~~ & ~~~~$B^{R+Z+M}$~~~~ & ~~~~~ \SN  \\[0.05cm]
       \hline
       \hline
       Local       & 2.5            & 0.57           & 0.50           & 0.24   \\[0.05cm]
       Equilateral & 6.7            & 4.7            & 4.3            & 0.018  \\[0.05cm]
       Orthogonal  & -5.1           & -1.13          & -1.05          & 0.037  \\[0.05cm]
			\hline
			\hline
			\SN          & 0.77           & 0.47           & 0.47           & ---    \\[0.05cm]
   \end{tabular}
   \caption{Correlations between the primordial templates and the intrinsic bispectra, computed as $\fnlcon = \F{B}{T}/\F{T}{T}$. The signal-to-noise \SN is given by the square root of the autocorrelation.}
\label{tab:fnl_results}
\end{table}

The most striking feature of \tref{tab:fnl_results} is the difference between the $B^R$ and $B^{R+Z}$ bispectra, with the former yielding a larger \fnl contamination. This is a quantitative confirmation of what shown in \fref{fig:analytical}, where the recombination-only curve exhibits a positive offset with respect to the integrated one showing the importance of the integrated effects which include $\Delta^{(1)}$. On the other hand, the time-integrated effects given by the metric affect $\fnlcon$ only marginally, and do not seem to affect the signal-to-noise. This can be seen by comparing the $B^{R+Z}$ and $B^{R+Z+M}$ columns of \tref{tab:fnl_results}.

The last column of \tref{tab:fnl_results} can be computed by using a first-order Boltzmann code. Our value of $\SN = 0.24$ for the local-template agrees with the one obtained using the first-order code CAMB \cite{lewis:2000a} and with Ref.~\cite{komatsu:2001a}.

Pitrou et al.~(2010) \cite{pitrou:2010a} found $\fnlcon \sim 5$ and $\SN(\lmax=2000) \sim 1$ by using the Boltzmann code \emph{CMBquick} \cite{pitrou:2011a}. In that code, the bispectrum was computed by including all \los sources in \eref{eq:line_of_sight_sources}, including lensing and time-delay, and integrating them until shortly after recombination. This is perfectly achievable since lensing and time-delay pose numerical problems only at later times, when small-scale multipoles get excited. However, the choice of the cutoff time is arbitrary as the time-integrated effects are important throughout cosmic evolution.
We ran \code with the same parameters and cutoff time as \emph{CMBquick}, and we obtained similar values: $\fnlcon = 3.7$ and $\SN(\lmax=2000) = 1.1$. As pointed out in \sref{sec:convergence}, the remaining discrepancy might be due to a lack of numerical convergence in \emph{CMBquick}. Furthermore, the most recent version of \emph{CMBquick} yields a value of $\fnlcon \sim 3$ \cite{PitrouPrivate} which is more in line with what we find.

\begin{figure}[t]
	\centering
		\includegraphics[width=0.7\linewidth]{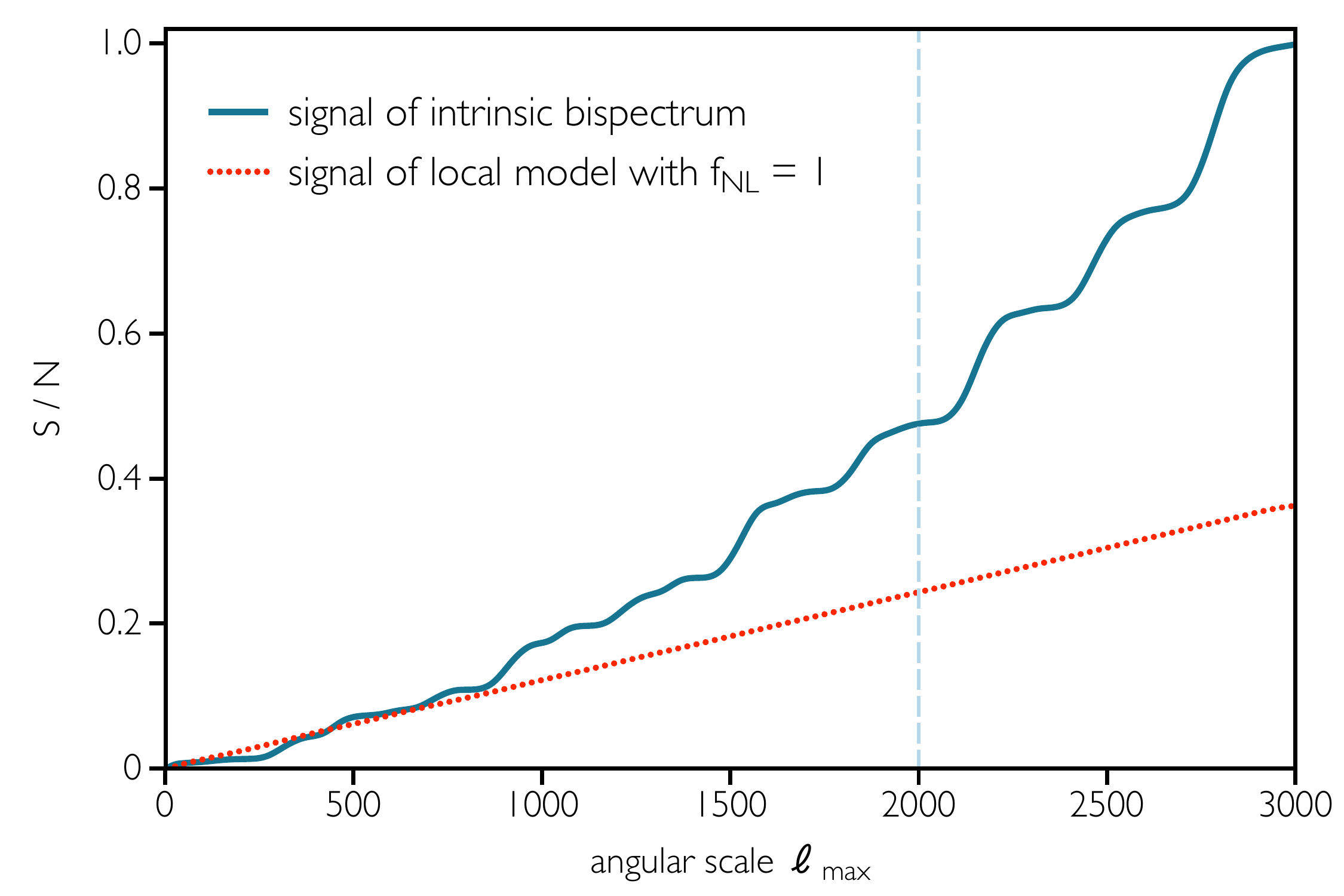}
	\caption{Signal-to-noise ratio of the $B^{R+Z+M}$ bispectrum, which includes all effects apart from time-delay and lensing. The S/N reaches unity at $\ell=3000$. At $\ell=2000$, the smallest scale probed by Planck, it is roughly equal to $0.5$.}
	\label{fig:signal_to_noise}
\end{figure}

In \fref{fig:signal_to_noise}, we show the signal-to-noise ratio of the $B^{R+Z+M}$ bispectrum as a function of \lmax, which is the angular resolution of the considered experiment. We find that, for Planck's resolution of $\lmax=2000$, the signal-to-noise ratio is roughly equal to $0.5$, and reaches unity only for $\lmax=3000$.

\section{Convergence tests}
\label{sec:convergence}

\begin{figure}[t]
	\centering
		\includegraphics[width=1\linewidth]{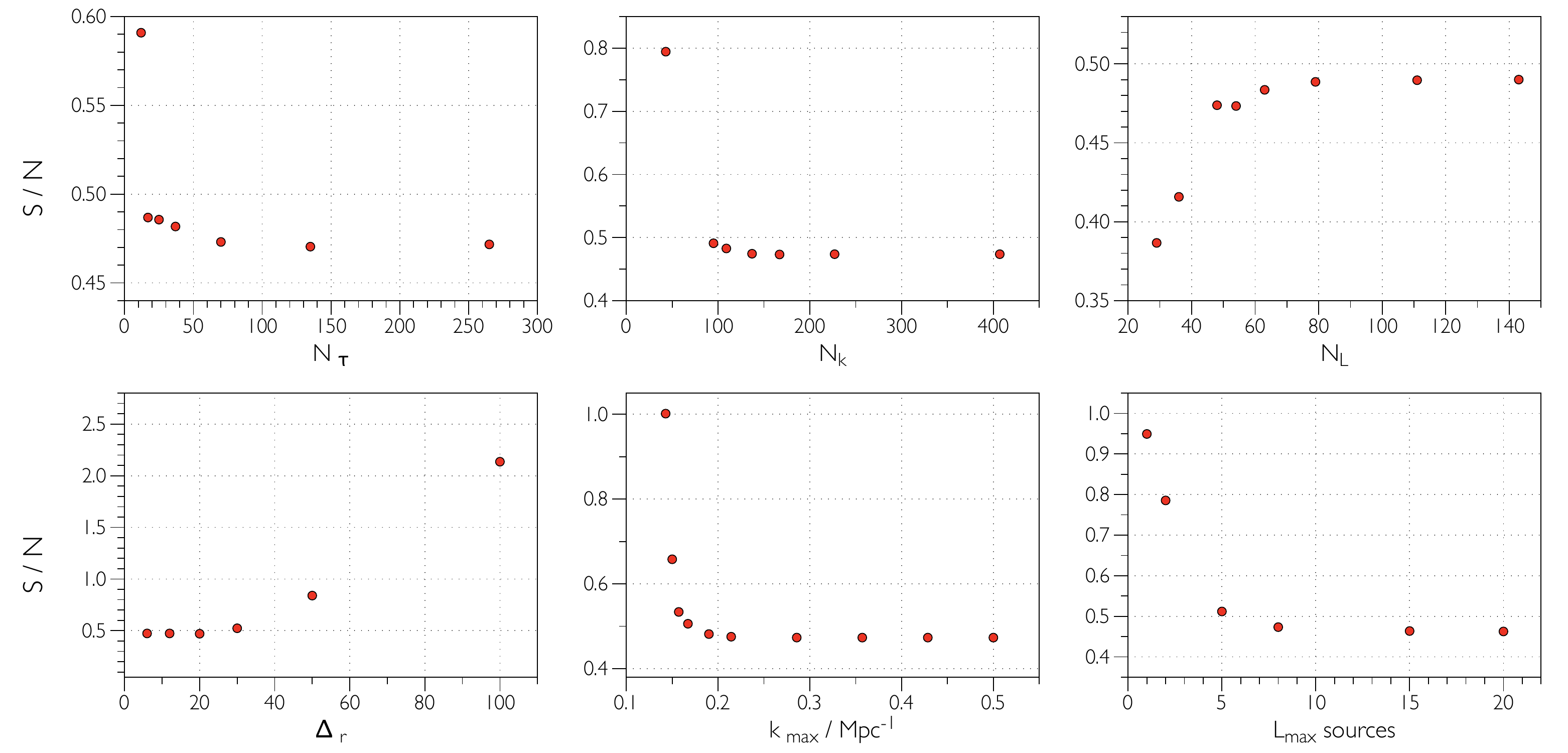}
	\caption{Convergence of the signal-to-noise ratio for $B^{R+Z+M}$, our most complete bispectrum. Refer to the text for details on the tested parameters.}
	\label{fig:convergence_tests}
\end{figure}

We checked the numerical robustness of our bispectrum results by varying the most relevant numerical parameters in \code:
\begin{itemize}
	\item $N_\tau$, number of sampling points in conformal time for the \los sources.
	\item $N_k$, number of sampling points per direction of k-space $(k,k_1,k_2)$ for the transfer functions.
	\item $N_L$, number of sampling points per direction of l-space $(l_1,l_2,l_3)$ for the bispectrum.
	\item $\Delta_r$, step size of the $r$-grid in the bispectrum integral in \eref{eq:bispectrum_integral}.
	\item $k_\text{max}$, maximum value of $k$ for which we compute the transfer functions.
	\item \Lmax, highest multipole source considered in the \los integral in \eref{eq:los_integral}.
\end{itemize}
In \fref{fig:convergence_tests}, we show how quickly the signal-to-noise of the intrinsic bispectrum converges for all the tested parameters. (Note that the convergence of $\fnlcon = \F{B}{T}/\F{T}{T}$ is even faster than the convergence of $\SN=\sqrt{\F{B}{B}}$ as numerical errors tend to cancel when taking ratios.)

We find that the signal strongly depends on the number of multipoles included in the \los integration, \Lmax, as shown in the bottom-right panel of \fref{fig:convergence_tests}. While at first order the \los sources consist only of a monopole, at second order the sum $j_{nm} \, S_{m}$ has to be cut at a suitable \Lmax ~-- see \eref{eq:los_integral}. We obtain a convergence only for $\Lmax>8$, with lower values yielding a larger signal. This behaviour might partly explain the large value of \fnlcon found by Pitrou et al.~(2010) \cite{pitrou:2010a}, who used $\Lmax=4$.

As mentioned in \sref{sec:bispectrum}, we compute the Fisher matrix elements in \eref{eq:estimator} by interpolating the bispectra on a mesh. The top-right panel of \fref{fig:convergence_tests} shows how our interpolation technique yields percent-level precision with just $60$ points out of $2000$ in each $\ell$-direction. We also tested the interpolation against known results, such as the signal-to-noise of the local model, and obtained the same level of agreement.

\section{Conclusions}
\label{sec:conclusions}

In this paper we have presented results from a new, efficient numerical code, \code, designed to calculate the CMB anisotropies up to second order.  We have exploited it to find the temperature bispectrum which arises even for purely Gaussian initial density perturbations. This intrinsic non-Gaussianity will necessarily bias attempts to estimate different types of primordial non-Gaussianity from the CMB bispectrum. 
The efficiency of \code has allowed us to demonstrate convergence of our results with respect to several different numerical parameters. We have also demonstrated per-cent-level agreement with analytical estimates in the squeezed limit, and we believe our answers are robust. 

The contamination from the intrinsic bispectrum generated by the second-order Einstein-Boltzmann equations generally leads to a small bias in the estimates of non-Gaussianity, which is good news for the prospect of using forthcoming data for the Planck satellite to probe primordial non-Gaussianity.  
While the precise answer depends on the terms included, the biases for local templates of non-Gaussianity are below the level of primordial \fnl detectable by Planck. The biases from the intrinsic bispectrum for other primordial templates, equilateral and orthogonal, also appear to be small.
The intrinsic non-Gaussianity can be searched for directly, using the predicted signal as a template; our calculations suggest this signal is just beyond what is possible with Planck, with a signal-to-noise rising to unity only for $\lmax=3000$ (\fref{fig:signal_to_noise}.) 

In comparing to recent calculations, we find good agreement with the results of Huang and Vernizzi \cite{huang:2012a} when we include the integrated redshift term with the recombination contribution.  The signal-to-noise for the intrinsic signal matches well, while our bias to \fnlcon $ \simeq 0.5$ is slightly different, which appears to be due to differences in the implementation of the local template. 
Excluding the integrated redshift term yields a significantly higher answer, with \fnlcon $= 2.5$.  This is much more similar to the results of Pitrou et al \cite{pitrou:2010a}, which  focussed on the contributions on the recombination surface alone.  We have also found that the number of multipole sources in the \los integral required for numerical convergence is $\Lmax\geq8$, and we find larger values of \fnlcon are obtained for $\Lmax=4$ as used in Ref.~\cite{pitrou:2010a}. Su et al.~\cite{su:2012a} find similar numerical values to Huang and Vernizzi \cite{huang:2012a} for the bias, but disagree on the signal-to-noise of the intrinsic signal.  We are unable to directly compare our numerical results with theirs, since they use integration by parts which leads to different \los source terms. We comment on this approach in \sref{subs:integration_by_parts}.

We have shown how the redshift terms along the \los lead to a change in the value of the local-type \fnlcon bias of approximately 2. We interpret this as the evidence that effects which are not at recombination are important, and should be all included in order to obtain a complete result. We plan to  further develop our numerical code to include the time-delay and lensing contributions. The time-delay effect was studied in Ref.~\cite{hu:2001a} and is expected to be small. The lensing term, on the other hand, is known to strongly correlate with the first-order ISW effect and thus yields a strong squeezed signal that contaminates the local measurement of Planck with a bias of $\fnlcon \sim 7$ \cite{lewis:2012a, hanson:2009a, smith:2011a, serra:2008a, lewis:2011a, lewis:2006a}.

We have calculated only the scalar ($m=0$) bispectrum, neglecting higher moments. This should give a reliable estimate of local-type \fnl since higher moments are suppressed for squeezed configurations. However there will be additional contributions to equilateral and orthogonal templates from higher moments and these still need to be evaluated.

It is our intention to make the bispectrum and Fisher parts of \code public later in 2013 \cite{pettinari:2013c} as independent modules for CLASS \cite{lesgourgues:2011a, blas:2011a}. We shall release the full code once the extensions discussed above have been completed.

\acknowledgments

We thank Nicola Bartolo, Martin Beneke, Alicia Bueno Belloso, Marco Bruni, Christian Byrnes, Paolo Creminelli, Dominic Galliano, Zhiqi Huang, Eiichiro Komatsu, Julien Lesgourgues, Donough Regan, Gianmassimo Tasinato, Thomas Tram, Filippo Vernizzi for useful discussions. We are particularly grateful to Antony Lewis for his insight on the the analytical approximation for the squeezed limit, to Cyril Pitrou for his help in comparing results, and to Leonidas Christodoulou for baptising our code, \code. 

RC, CF, KK and DW gratefully acknowledge financial support from the STFC grant ST/H002774/1. KK also thanks the ERC and the Leverhulme trust. GWP is supported by the Leverhulme trust.

\clearpage
\bibliographystyle{JHEP}
\bibliography{fnl_paper}

\end{document}